\documentclass[aps,pra,twocolumn]{revtex4-1}
\usepackage{amsmath,amssymb}
\usepackage{graphicx}
\usepackage[caption=false]{subfig}
\usepackage{mathtools}
\usepackage[pdftex,colorlinks=true,allcolors=blue]{hyperref}
\newcommand{\Rt}{\widetilde{R}}
\newcommand{\RtI}{\widetilde{R}_1}
\newcommand{\RtII}{\widetilde{R}_2}
\DeclarePairedDelimiter\floor{\lfloor}{\rfloor}

\begin{document}

\title{Discontinuities in the Electromagnetic Fields of Vortex Beams in the Complex Source/Sink Model}

\author{Andrew Vikartofsky, Liang-Wen Pi, and Anthony F. Starace}

\affiliation{Department of Physics and Astronomy, The University of Nebraska, Lincoln, Nebraska 68588-0299, USA}

\date{March 7, 2017}

\begin{abstract}
	An analytical discontinuity is reported in what was thought to be the discontinuity-free exact nonparaxial vortex beam phasor obtained within the complex source/sink model.  This discontinuity appears for all odd values of the orbital angular momentum mode.  Such discontinuities in the phasor lead to nonphysical discontinuities in the real electromagnetic field components.  We identify the source of the discontinuities, and provide graphical evidence of the discontinuous real electric fields for the first and third orbital angular momentum modes.  A simple means of avoiding these discontinuities is presented.
\end{abstract}

\maketitle

\section{Introduction}

The worldwide effort to develop increasingly powerful lasers will allow the exploration of new physical regimes of intense laser interactions with matter as well as the development of new applications that such intense laser regimes permit~\cite{Mourou2006,DiPiazza2012}. Experimentally, the highest laser intensities are obtained using tight focusing techniques, in which the laser spot size in the focal region is comparable to the laser field wavelength. Theoretical simulations of laser-matter interactions under such tight focusing conditions require a detailed description of the laser fields in the focal region that goes beyond the paraxial approximation~\cite{Lax1975,Barton1989,Sepke2006,Sepke2006-2,Hu2006,Salamin2007,Pi2015}.

Laser beams that carry orbital angular momentum (OAM) provide another means of investigation into laser-matter interactions~\cite{Torres2011,Yao2011,Andrews2013}.  Consideration of light with nonzero OAM has been increasing in many fields including harmonic generation~\cite{Zurch2012,Hernandez-Garcia2013,Gariepy2014}, particle acceleration~\cite{Lembessis2016,Vaziri2015}, and quantum information~\cite{Molina2007,Xiao:16}.  Multiple nonparaxial analytic representations~\cite{Richards1959,Caron1999,Bandres2004,Lin2006} have been developed to model tightly focused beams with nonzero OAM.  One particularly important representation for a free space beam is the Laguerre-Gaussian (LG) basis, which can be used to represent optical vortices of any angular momentum mode~\cite{Siegman}.

The \textit{complex point-source} model~\cite{Deschamps,Shin:77,Cullen1979} is one tool that has been developed to analytically describe focused beams carrying OAM.  This model is used to find solutions to the nonparaxial Helmholtz equation.  This model assumes that the beam source exists at a complex point whose real value lies along the beam's axis, and that the beam can be represented by an outgoing spherical wave.  It was shown by M. Couture and P. A. Belanger~\cite{Couture1981} that (for an appropriate choice of boundary conditions) the spherical waves represented by this model are equivalent to the paraxial representation of a Gaussian (zero OAM) beam with all perturbative corrections included.  The major benefit of this method is that it provides a closed form analytical representation of the beam's phasor, which is the complex function of the beam's spatiotemporal amplitude and phase that satisfies the scalar Helmholtz equation~\cite{Siegman,Andrews2013}.  This is a distinct advantage of the complex point-source model as compared to other methods~\cite{Richards1959,Caron1999,Bandres2004}, in which the fields are usually defined using either a series or an integral representation.  The complex point-source model, however, still has one major drawback.  Namely, the point-source Gaussian phasor solution is known to contain singularities in its square modulus as well as a discontinuity at the beam waist~\cite{Ulanowski:00}.

The \textit{complex source/sink} model~\cite{Ulanowski:00} was developed to avoid the discontinuity and singularities encountered in the complex point-source model.  The complex source/sink model represents the beam as two counter-propagating spherical waves, both centered at the imaginary location used in the complex point-source model.  In this new model, the singularities and discontinuity in the square modulus of the Gaussian phasor both vanish.

In this paper, we show that the discontinuities still arise in phasors generated from the complex source/sink model for all odd OAM modes.  The discontinuity in the phasor leads directly to discontinuities in the electromagnetic (EM) fields.  Thus, real fields generated from the complex source/sink phasor are nonphysical for odd OAM values.

This paper is organized as follows.  In Section~\ref{sec:methods}, we discuss use of the phasor in determining the EM fields and highlight the source of their discontinuities.  In Section~\ref{sec:DCphasor}, we demonstrate analytically why the discontinuity appears in the phasor for odd OAM, and why it does not appear for even OAM.  It is also shown how the discontinuity can be avoided.  In Section~\ref{sec:DCfields}, we present numerical results illustrating the discontinuities in electric field components that result from the discontinuity in the phasor.  In Section~\ref{sec:conc}, we summarize our results and present our conclusions.

\section{The phasor and field equations}\label{sec:methods}

Traditionally, solving the full Helmholtz problem involves finding six field solutions to the vector Helmholtz equations.  Matters are greatly simplified when instead one needs to find only a single solution to the scalar Helmholtz equation.  This one solution is the beam's phasor.

From a general expression for a phasor, Hertz potentials~\cite{Stratton,Jackson} (alternatively ``Hertz vectors" or ``polarization potentials") can be used to generate exact expressions for the complex EM fields.  The Hertz vectors, defined in Eq.~\eqref{E:Hertz} for a linearly polarized beam propagating in the $\mathbf{\hat{z}}$-direction, are represented in general as the complex phasor with a direction chosen based on the beam polarization:

\begin{subequations}\label{E:Hertz}
	\begin{align}
	\mathbf{\Pi}_e=&~\psi(\mathbf{r},t)\,\mathbf{\hat{x}}\\
	\mathbf{\Pi}_m=&~\eta_0\,\psi(\mathbf{r},t)\,\mathbf{\hat{y}}
	\end{align}
\end{subequations}

\noindent Here, $\psi(\mathbf{r},t)$ is the phasor and $\eta_0$ is the impedance of free space.

The Hertz potentials are sometimes referred to as ``super potentials" because they directly generate the usual scalar and vector EM potentials, which in turn generate the EM fields.  Consequently, the complex vector fields $\textbf{E}$ and $\textbf{B}$ can be obtained directly from the Hertz potentials~\cite{Stratton}, and therefore from the phasor.

\begin{subequations}\label{E:hertzfields}
	\begin{align}
	\mathbf{E}=&~\nabla\times\nabla\times\mathbf{\Pi}_e-\mu_0\frac{\partial}{\partial t}\left(\nabla\times\mathbf{\Pi}_m\right)\\
	\mathbf{H}=&~\nabla\times\nabla\times\mathbf{\Pi}_m+\epsilon_0\frac{\partial}{\partial t}\left(\nabla\times\mathbf{\Pi}_e\right)
	\end{align}
\end{subequations}

Using the complex source/sink model, April~\cite{April2008,April2010} proposed an analytically exact discontinuity-free representation of the phasor for nonparaxial LG beams of any radial and OAM mode.  April's methods have since been adopted in many other works (e.g.,~\cite{Marceau2012,Chu2012,Marceau2013,Sell2013,Fillion2015,Marceau2015,Wong2015,Varin2016}).  As long as one considers only the square modulus of phasor solutions derived from the complex source/sink method, such as April's phasor, the discontinuity and singularities are absent as claimed~\cite{April2010,Sheppard1998}.  This does not mean, however, that the phasors themselves are discontinuity free.  As we show in Section~\ref{sec:DCphasor}, consideration of the real and/or imaginary parts of the source/sink phasor, depending on the choice of initial phase $\phi_0$, very clearly reveals a discontinuity at the beam waist for certain parameters.  The presence of this axial discontinuity depends on the choice between two representations of the complex radius of curvature of the spherical waves, $\Rt$~\cite{April2010,Sheppard:07}.

Most work using April's phasor~(e.g.,~\cite{Marceau2012,Chu2012,Marceau2013,Sell2013,Marceau2015,Varin2016}) has so far been done with the lowest order LG mode (the ``Gaussian mode,''  which has zero OAM) or by considering the phasor only in the paraxial limit.  As we will show, the phasors for these two common cases are not affected by this discontinuity.

\section{Discontinuity in the Phasor}\label{sec:DCphasor}

April~\cite{April2010} combined the complex source/sink method with use of a Poisson-like frequency spectrum~\cite{Caron1999,Feng2000}, ${f(\omega)}$, to analytically represent the generic phasor ${U_{p,n}}$ from which EM fields can be derived using the Hertz potentials.  For the zero order radial mode (${p=0}$), April's phasor for the nonparaxial LG beam with any OAM index $n$ can be expressed as (see Eqs.~(16)~\&~(17) of \cite{April2010})

\begin{equation}\label{E:fullPhasor}
\begin{aligned}
	U_{0,n}(\mathbf{r},\omega)=&~\frac{4\cos(n\phi)}{(2n-1)!!}f(\omega)\left(\frac{ka}{2}\right)^{1+n/2}\\
	\times&~\exp(-ka)P^n_n(\chi)j_n(k\Rt),
\end{aligned}
\end{equation}

\noindent where $j_n$ is the spherical Bessel function, $a$ is the confocal parameter of the focused beam, $\phi$ is the cylindrical angle, and the complex-valued associated Legendre function $P_n^n(\chi)$ is defined by Eqs.~(8.6.6) and~(8.6.18) of Ref.~\cite{Abramowitz},

\begin{equation}\label{E:Legendre}
	P_n^n(\chi)=\left(\chi^2-1\right)^{n/2}\frac{d^n}{d\chi^n}\left(\frac{1}{2^n n!}\frac{d^n\left(\chi^2-1\right)^n}{d\chi^n}\right),
\end{equation}
in which the complex argument, $\chi$, is defined by

\begin{equation}\label{E:chi}
	\chi \equiv (z+ia)/\Rt.
\end{equation}

\noindent 
\noindent There are two choices (cf.~Eq.~(14) of Ref.~\cite{April2010}) for the complex spherical radius of curvature, $\Rt$, in Eq.~\eqref{E:fullPhasor}:

\begin{subequations}\label{E:Rts}
\begin{align}
	\RtI=&\sqrt{\rho^2+(z+ia)^2}\label{E:Rts-a}\\
	\RtII=&i\sqrt{-\rho^2-(z+ia)^2}\label{E:Rts-b},
\end{align}
\end{subequations}

\noindent where ${\rho,\phi,z}$ are the cylindrical coordinates in which $\mathbf{\hat{z}}$ is the direction of propagation.  The Poisson-like frequency spectrum ${f(\omega)}$ in Eq.~\eqref{E:fullPhasor} is defined as (see Eq.~(4) of~\cite{Caron1999} or Eq.~(20) of~\cite{April2010})

\begin{equation}\label{E:PoissonSpectrum}
f(\omega)=2\pi e^{i\phi_0} \left(\frac{s}{\omega_0}\right)^{s+1}\frac{\omega^s\exp(-s\omega/\omega_0)}{\Gamma(s+1)}\theta(\omega),
\end{equation}

\noindent where $s$ is the spectral parameter~\cite{Caron1999,Feng2000} (which is related to the bandwidth of the pulse, which in turn is related to its duration), $\omega_0$ is the frequency at which $f(\omega)$ has its maximum, $\phi_0$ is the phase of the pulse, and ${\theta(\omega)}$ is the Heaviside unit step function.  

It has been stated~\cite{April2010, Sheppard:07} that neither choice of $\Rt$ in Eq.~\eqref{E:Rts} would cause the phasor to suffer from discontinuities, but we will show that only the choice $\RtII$ produces continuous phasor components across the beam waist for all values of OAM.

Note also that the associated Legendre functions defined in Eq.~\eqref{E:Legendre} contain a branch cut only for odd index $n$.  The following sections will elucidate the interplay between this branch cut and the choice of $\Rt$, and show how this determines whether or not the phasors contain discontinuities.

\subsection{Odd OAM Modes}\label{sec:odds}

Inspection of Eqs.~\eqref{E:fullPhasor}-\eqref{E:PoissonSpectrum} shows that only the last two factors in the phasor may lead to the existence of a discontinuity. We thus focus on these two factors and express Eq.~\eqref{E:fullPhasor} as 

\begin{equation}\label{E:phasorDCParts}
	U_{0,n}(\mathbf{r}, \omega) = c_n(\phi, \omega) P_n^n(\chi) j_n(k\Rt),
\end{equation}
where $c_n(\phi, \omega)$ is defined by comparison of Eqs.~\eqref{E:fullPhasor} and~\eqref{E:phasorDCParts}.
To illustrate how the choice of $\Rt$ determines whether or not there is a discontinuity in the phasor, we consider the simplest odd OAM mode, ${n=1}$.  We first use the choice $\RtI$ to demonstrate a discontinuity at the beam waist, ${z=0}$.

\subsubsection{Exact expansion of $U_{0,1}$ in powers of $\Rt$}
 
Expressing the spherical Bessel function in Eq.~\eqref{E:phasorDCParts} in terms of sines and cosines [cf.~Eqs.~\eqref{E:A:jn}--\eqref{E:A:Q}] and defining the parameter

\begin{equation}\label{E:xi}
	\xi\equiv k\Rt,
\end{equation}

\noindent the ${n=1}$ phasor may be expressed as

\begin{equation}
	U_{0,1} = c_1P^1_1(\chi) \left( -\frac{\cos(\xi)}{\xi} + \frac{\sin(\xi)}{\xi^2} \right).
\end{equation}

\noindent Replacing the trigonometric functions by their series expansions, we obtain

\begin{equation}
\begin{aligned}
	U_{0,1}=c_1 P^1_1(\chi) \left[ -\frac{1}{\xi}\sum_{m=0}^\infty\frac{(-1)^m \xi^{2m}}{(2m)!} \right.\\
	\left.+\frac{1}{\xi^2}\sum_{m=0}^\infty\frac{(-1)^m \xi^{2m+1}}{(2m+1)!} \right].
\end{aligned}
\end{equation}

\noindent Combining the two summations, we obtain:

\begin{subequations}\label{E:U01 in powers of R}
\begin{align}
	U_{0,1}=&~c_1 P^1_1(\chi) \frac{1}{\xi}\sum_{m=0}^\infty \kappa_m\,\xi^{2m}\label{E:U01 in powers of R-a}\\
	\kappa_m\equiv&~(-1)^{m+1}\frac{2m}{(2m+1)!}
\end{align}
\end{subequations}

\noindent where, from Eq.~\eqref{E:Legendre},

\begin{equation}\label{P11}
	P^1_1(\chi)=\sqrt{\chi^2-1}.
\end{equation}

\subsubsection{$U_{0,1}$ with the choice $\Rt=\RtI$}

Making the choice ${\Rt=\RtI}$ [defined in Eq.~\eqref{E:Rts-a}] in Eqs.~\eqref{E:chi} and~\eqref{E:xi},  $U_{0,1}$ in Eq.~\eqref{E:U01 in powers of R-a} becomes:
\begin{equation}\label{E:m1Rt1}
\begin{aligned}
	U_{0,1}= &\sqrt{\frac{-\rho^2}{\rho^2+(z+ia)^2}} \cdot \frac{1}{\sqrt{\rho^2+(z+ia)^2}}\\
	&\times c_1 \sum_{m=0}^\infty(\kappa_m k^{2m-1}) \left(\rho^2+(z+ia)^2\right)^{m}.
\end{aligned}
\end{equation}

\noindent We see that the summation in Eq.~\eqref{E:m1Rt1} involves integer powers of complex numbers, whereas the prefactors multiplying the summation include two square roots of complex numbers, whose evaluation requires some care. In general, when dealing with products of square roots of complex numbers, it is best to evaluate each square root separately by expressing each complex number in terms of its magnitude and phase before taking its square root. In particular, multiplying the arguments of two square roots before taking the square root can lead to erroneous results. (For example, $\sqrt{-1}\cdot\sqrt{-1}=i\cdot i=-1$, but $\sqrt{-1\cdot-1}=\sqrt{1}=1$.) Thus, we have expressed each of the complex arguments of the two square root prefactors in Eq.~\eqref{E:m1Rt1} in polar notation before taking the square roots.  The result is:
\begin{subequations}
\begin{align}
	U_{0,1}=&~c_1\exp\left(\frac{i}{2}\left(\phi_1-\phi_2\right)\right) \sum_{m=0}^\infty\lambda_m\exp(im\phi_2)\label{E:R1twoExp}\\
	\phi_1=&~\arctan\left(\frac{2az}{-\rho^2+a^2-z^2}\right) \label{E:R1twoExp-b}\\ 
	\phi_2=&~\arctan\left(\frac{2az}{\rho^2-a^2+z^2}\right) \label{E:R1twoExp-c} \\
	\nonumber \lambda_m\equiv&~(\kappa_m k^{2m-1})\rho\\
	 &\times \left[(\rho^2+z^2+a^2)^2-(2a\rho)^2\right]^{(m-1)/2}\label{lambda-m}.
\end{align}
\end{subequations}
\noindent Here, the real numbers $\lambda_m$ are $m$-dependent magnitudes, defined in Eq.~\eqref{lambda-m}, and $\phi_1$ and $\phi_2$ are the phases of the complex numbers inside the first and second square root prefactors in Eq.~\eqref{E:m1Rt1} (which originate from $P_1^1(\chi)$ and $\RtI$ respectively).  The $\arctan$ function is defined over ${-\pi<\phi\le\pi}$; thus, $\arctan$ has a branch cut along the negative real axis. At the beam waist ${z=}0$, the imaginary parts of the complex numbers whose phases are given by $\phi_1$ and $\phi_2$ are zero; thus, the branch cut along the negative real axis of each $\arctan$ function in Eqs.~\eqref{E:R1twoExp-b} and~\eqref{E:R1twoExp-c} is determined by the region over which the denominators in each of their arguments is negative.  At $z=0$ the denominator of $\phi_1$ is negative for $\rho>a$, while that for $\phi_2$ is negative for $\rho<a$. 

\begin{figure}[h!]%
	\centering
		\label{fig:phi1}%
		\includegraphics[height=2.4in]{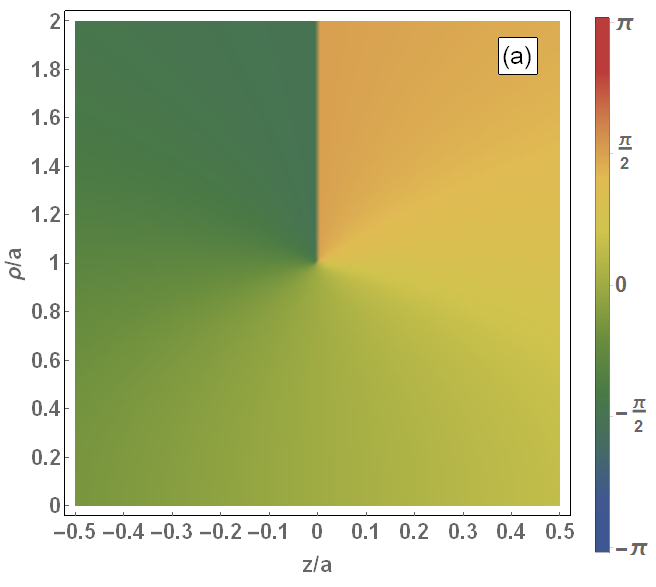}
	\\
		\label{fig:phi2}%
		\includegraphics[height=2.4in]{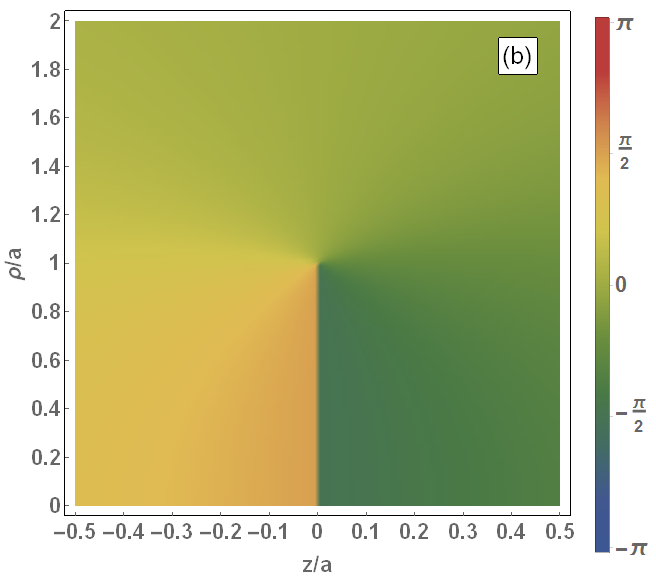}
	\\
		\label{fig:phi12}%
		\includegraphics[height=2.4in]{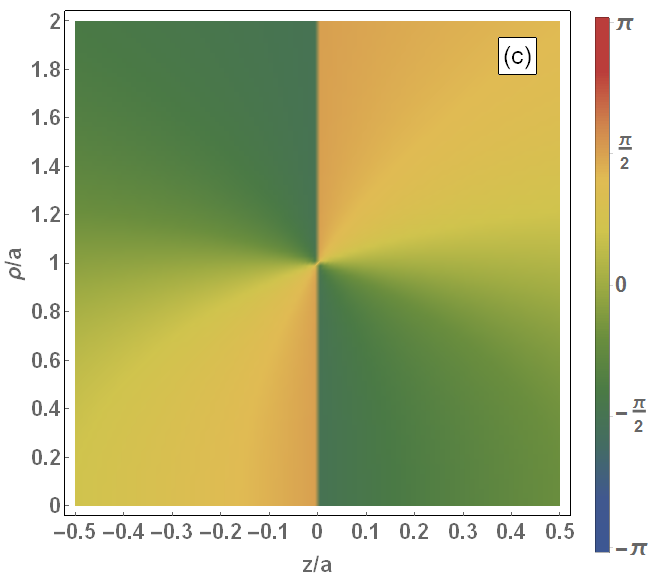}
	\caption[]{The phases (a) ${\phi_1/2}$, (b) ${-\phi_2/2}$, and (c) ${(\phi_1-\phi_2)/2}$ as functions of $\rho/a$ and $z/a$, where $a$ is the confocal parameter of the focused laser beam. Values of each phase over the range from $-\pi$ to $+\pi$ are indicated by the vertical color coding strip to the right of each panel.  A phase jump of $\pi$ occurs for $\rho/a>1$ in (a), for $\rho/a<1$ in (b), and for all values of $\rho/a$ in (c). See text for discussion.}
	\label{fig:phases}%
\end{figure}

The $\phi_1$ and $\phi_2$ phase factors multiplying the sum in Eq.~\eqref{E:R1twoExp} always have a phase difference of $\pi$ across the branch cut due to their overall factor of $1/2$ in the exponential. The key point is that $\phi_1$ and $\phi_2$ have branch cuts over different regions of the parameter $\rho/a$.  Specifically,  $U_{0,1}$ is discontinuous for $\rho>a$ at $z=0$ owing to the change in sign of $\phi_1/2$ across the branch cut, while for $\rho<a$ it is discontinuous owing to the change in sign of $\phi_2/2$ across the branch cut.  Consequently, $U_{0,1}$ is discontinuous across the beam waist at $z=0$ for all values of $\rho/a$ owing to the discontinuity in the product of phases, $\exp\left(\frac{i}{2}\left(\phi_1-\phi_2\right)\right)$. These ranges of the ratio $\rho/a$ over which the discontinuities in the phases $\phi_1/2$, $-\phi_2/2$, and $\left(\phi_1-\phi_2\right)/2$ occur are illustrated in the three panels of Fig.~\ref{fig:phases}.  

Note that for each term in the sum in Eq.~\eqref{E:R1twoExp}, there is a phase factor involving an integer multiple of $\phi_2$.  However, each of these terms is continuous across the branch cut since each branch contains an integer number $m$ of full periods, resulting in a $2\pi$ phase difference across the branch cut.  Thus, the terms in the sum do not contribute to any discontinuity.

\subsubsection{$U_{0,1}$ with the choice $\Rt=\RtII$}

\begin{figure}[h!]
	\centering
	\subfloat[][]{%
		\label{fig:phi3}%
		\includegraphics[height=2.42in]{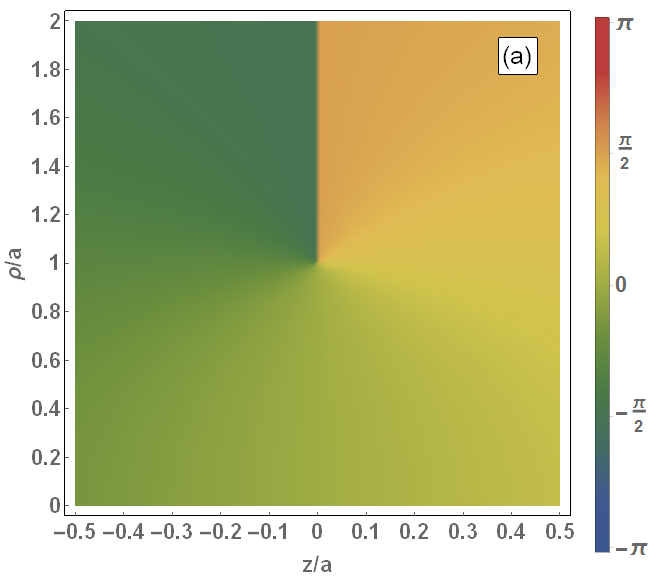}}%
	\\
	\subfloat[][]{%
		\label{fig:phi13}%
		\includegraphics[height=2.42in]{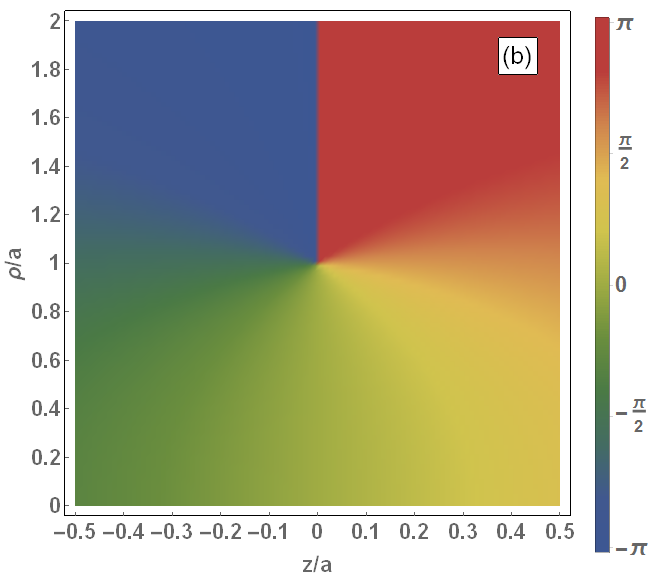}}%
	\caption[]{The phases (a) ${-\phi_3/2}$ and (b) ${(\phi_1-\phi_3)/2}$ as functions of $\rho/a$ and $z/a$, where $a$ is the confocal parameter of the focused laser beam and the behavior of the phase $\phi_1/2$ is shown in Fig.~\ref{fig:phases}(a). Values of each phase over the range from $-\pi$ to $+\pi$ are indicated by the vertical color coding strip to the right of each panel.  For $\rho/a>1$ a phase jump of $\pi$ occurs in (a) and a phase jump of $2\pi$ occurs in (b). See text for discussion.}
	\label{fig:phase products}%
\end{figure}

Use of the choice ${\Rt=\RtII}$ results instead in the phasor $U_{0,1}$ being continuous, as may be seen using the same arguments as in the previous section.  Specifically,  we replace $\RtI$ by $\RtII$[defined in Eq.~\eqref{E:Rts-b}] in Eqs.~\eqref{E:chi} and~\eqref{E:xi} and substitute the results in Eq.~\eqref{E:U01 in powers of R-a}.  Since ${\RtI^2=\RtII^2}$, the terms in the summation are continuous across the branch cut.  We thus focus on the new square root prefactors (corresponding to those for ${\Rt=\RtI}$ in Eq.~\eqref{E:m1Rt1}):

\begin{equation}\label{E:m1Rt2}
	U_{0,1}\propto \sqrt{\frac{-\rho^2}{\rho^2+(z+ia)^2}} \cdot \frac{1}{\sqrt{-\rho^2-(z+ia)^2} }.
\end{equation}

\noindent The number inside the first square root factor is the same as in Eq.~\eqref{E:m1Rt1}; consequently, it has the same phase factor, $\exp(i\phi_1)$. The number inside the square root in the denominator of the second factor in Eq.~\eqref{E:m1Rt2}  has the phase factor $\exp(i\phi_3)$, where
\begin{equation}\label{phi3}
\phi_3=\arctan\left(\frac{-2az}{-\rho^2+a^2-z^2}\right).
\end{equation}

\noindent Thus, the phasor has the same form as in Eq.~\eqref{E:R1twoExp}, but with a different phase outside the sum, i.e., 

\begin{equation}
U_{0,1}=-ic_1\exp\left(\frac{i}{2}\left(\phi_1-\phi_3\right)\right) \sum_{m=0}^\infty\lambda_m\exp(im\phi_2)\label{E:R2twoExp}
\end{equation}

By considering the branch cut in $\arctan$, one can see that both $\phi_1$ and $\phi_3$ are discontinuous in the same region, namely for $\rho>a$.  In both cases, the value changes sign as the $z=0$ plane is crossed.  When these two phase factors are multiplied together as in Eq.~\eqref{E:R2twoExp}, each one has a phase jump of $\pi$ (cf. Figs.~\ref{fig:phases}(a) and~\ref{fig:phase products}(a)), so that their product has a phase jump of $2\pi$, as shown in Fig.~\ref{fig:phase products}(b).  Hence, the phasor defined by Eq.~\eqref{E:R2twoExp} is continuous across the branch cut.

\subsubsection{Case of Arbitrary Odd OAM Modes}

We may easily see that for any odd OAM index $n$ in Eq.~\eqref{E:phasorDCParts}, the phasor $U_{0,n}$ will exhibit the same behaviors as just shown for the $n=1$ case. First, the associated Legendre function $P_n^n(\chi)$ in Eq.~\eqref{E:Legendre} always introduces a square root factor as on the right hand side of Eq.~\eqref{P11} for any odd index $n$, which in turn results in the first square root factor in Eqs.~\eqref{E:m1Rt1} and~\eqref{E:m1Rt2} regardless of whether one chooses respectively $\Rt=\RtI$ or $\Rt=\RtII$.  Second, the spherical Bessel function factor $j_n$ in Eq.~\eqref{E:phasorDCParts} will always introduce the second square root factor in Eqs.~\eqref{E:m1Rt1} and~\eqref{E:m1Rt2}, depending respectively upon whether one chooses $\Rt=\RtI$ or $\Rt=\RtII$.  One may see this by examining the expression for the spherical Bessel function given in Eq.~\eqref{E:A:jn}.  Specifically, for odd $n$ the square root factor comes from the factor $1/\Rt$ outside the square brackets in Eq.~\eqref{E:A:jn}; for odd $n$ the two summations inside the square brackets in Eq.~\eqref{E:A:jn} involve only even powers of $\Rt$ and hence do not contribute any square root factors.  Thus, the discontinuity in the phasor $U_{0,n}$ for a particular choice of $\Rt$ has the same behavior for any odd OAM $n$.

\subsection{Even OAM Modes}\label{sec:evens}

\begin{figure*}[t!]
	\subfloat[][]{%
		\label{fig:phi0mod2}%
		\includegraphics[height=1.9in]{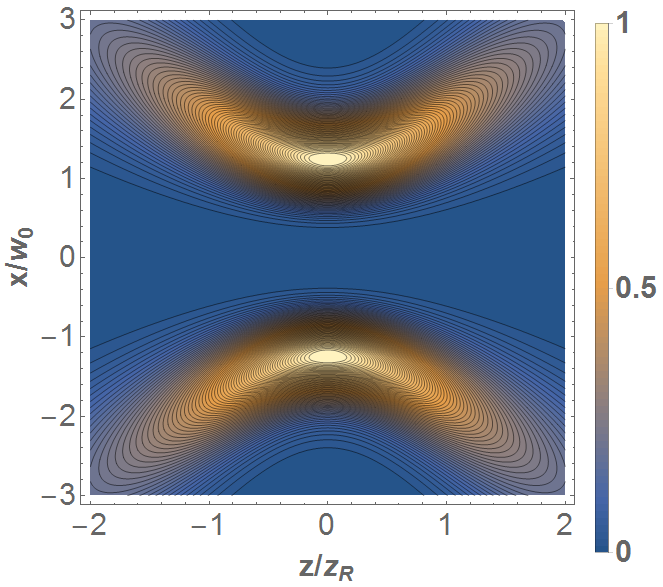}}
	\hspace{8pt}%
	\subfloat[][]{%
		\label{fig:phi0real}%
		\includegraphics[height=1.9in]{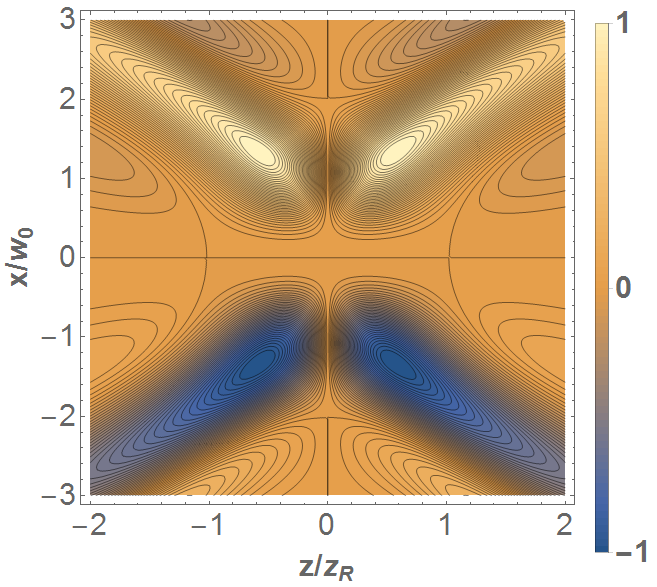}}
	\hspace{8pt}%
	\subfloat[][]{%
		\label{fig:phi0imag}%
		\includegraphics[height=1.9in]{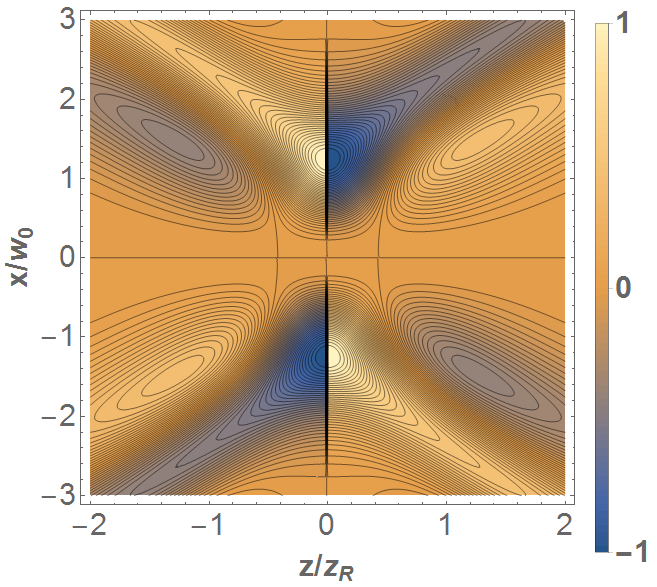}}
	\\
	\subfloat[][]{%
		\label{fig:phi0p25pimod2}%
		\includegraphics[height=1.9in]{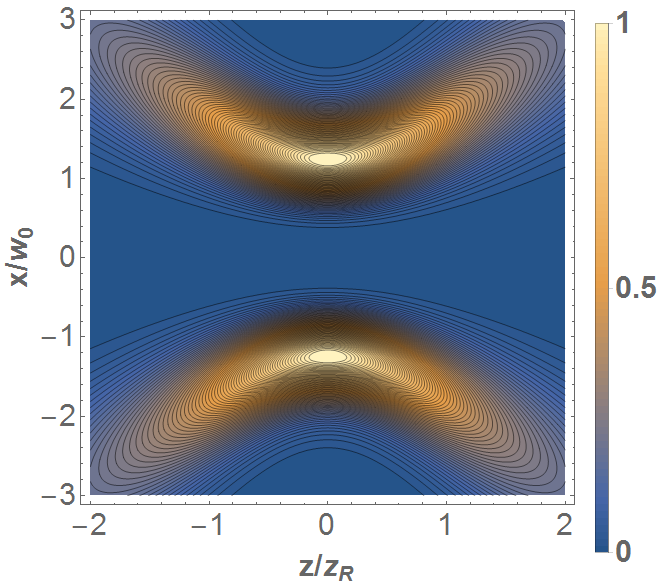}}
	\hspace{8pt}%
	\subfloat[][]{%
		\label{fig:phi0p25pireal}%
		\includegraphics[height=1.9in]{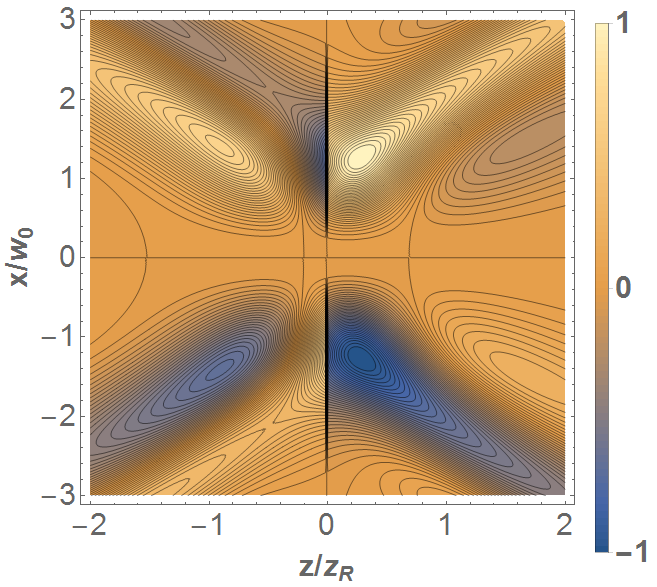}}
	\hspace{8pt}%
	\subfloat[][]{%
		\label{fig:phi0p25piimag}%
		\includegraphics[height=1.9in]{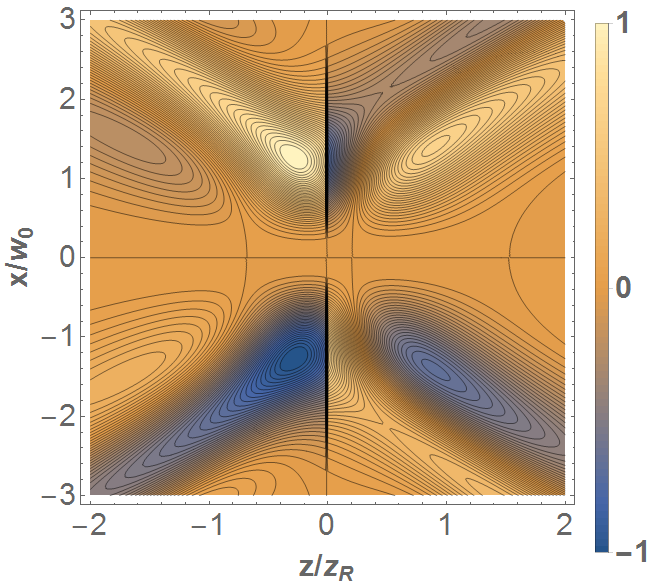}}
	\\
	\subfloat[][]{%
		\label{fig:phi0p5pimod2}%
		\includegraphics[height=1.9in]{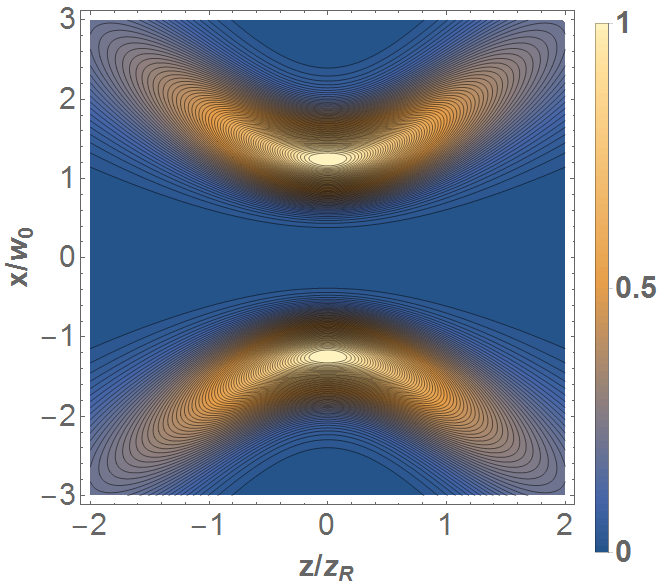}}
	\hspace{8pt}%
	\subfloat[][]{%
		\label{fig:phi0p5pireal}%
		\includegraphics[height=1.9in]{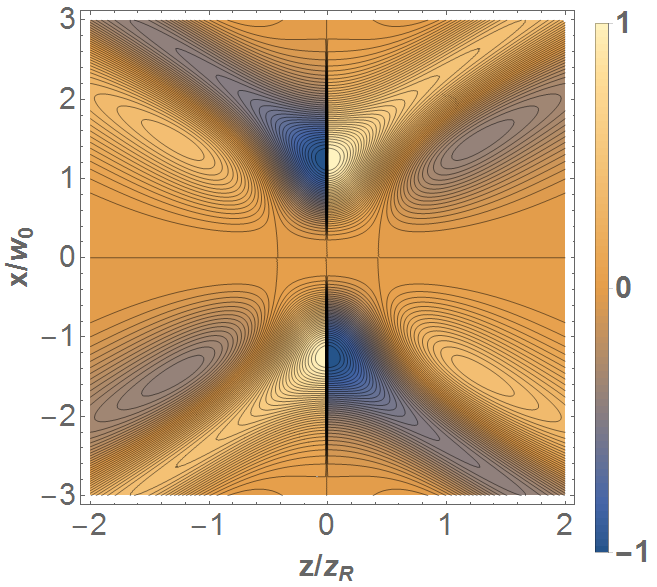}}
	\hspace{8pt}%
	\subfloat[][]{%
		\label{fig:phi0p5piimag}%
		\includegraphics[height=1.9in]{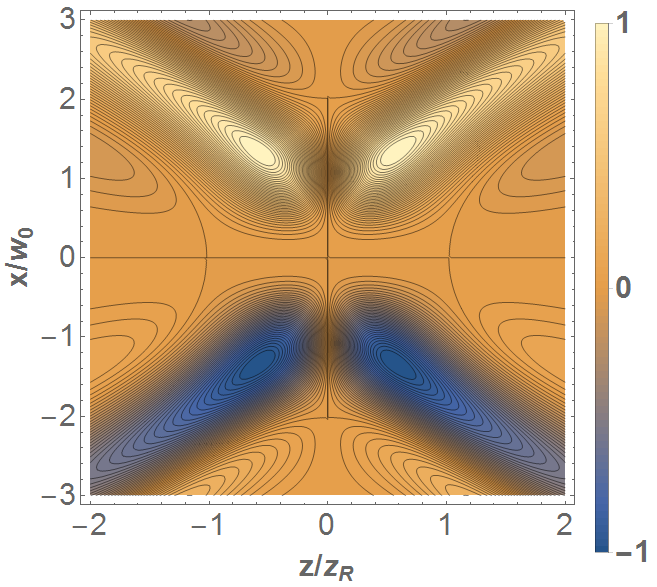}}
	
	\caption[]{The square modulus [\subref{fig:phi0mod2},\subref{fig:phi0p25pimod2},\subref{fig:phi0p5pimod2}], real part [\subref{fig:phi0real},\subref{fig:phi0p25pireal},\subref{fig:phi0p5pireal}], and imaginary part [\subref{fig:phi0imag},\subref{fig:phi0p25piimag},\subref{fig:phi0p5piimag}] of the phasor $U_{0,n}(\mathbf{r},t)$ in Eq.~\eqref{E:Fourier} for n=3 for phases ${\phi_0=0}$ [\subref{fig:phi0mod2}-\subref{fig:phi0imag}], $\phi_0=\pi/4$ [\subref{fig:phi0p25pimod2},\subref{fig:phi0p25pireal},\subref{fig:phi0p25piimag}], and ${\phi_0=\pi/2}$ [\subref{fig:phi0p5pimod2},\subref{fig:phi0p5pireal},\subref{fig:phi0p5piimag}].  Here, ${x,y,z}$ are the Cartesian coordinates.  The real and imaginary parts of the phasor are normalized to have a maximum amplitude of unity, and were calculated using the choice $\Rt=\RtI$ at ${y=0}$ and ${t=z/c}$.  The linearly polarized beam is assumed to have a spectral parameter ${s=712}$, beam waist ${w_0=2~\mu m}$, wavelength ${\lambda=800~nm}$, and Rayleigh length ${z_R\approx15.7\mu m}$. See text for discussion.}
	\label{fig:phasors}%
\end{figure*}

For even OAM modes $n$, the general expression for the phasor in Eq.~\eqref{E:fullPhasor} has the same form as in Eq.~\eqref{E:phasorDCParts}.  As has already been noted above, the associated Legendre function defined in Eq.~\eqref{E:Legendre} does not have a branch cut for even index $n$. We thus focus on the spherical Bessel function $j_n$ in Eq.~\eqref{E:phasorDCParts}, using the expression for $j_n$ in Eq.~\eqref{E:A:jn}.  From Eqs.~\eqref{E:A:P} and~\eqref{E:A:Q}  we see that for any OAM mode $n$ the functions $P$ and $Q$ involve respectively even and odd powers of $\Rt$.  For even $n$, the sine and cosine functions in Eq.~\eqref{E:A:jn} may be expanded respectively in terms of odd and even powers of $\Rt$.  Thus the two terms inside the square bracket in Eq.~\eqref{E:A:jn} each involve odd powers of $\Rt$.  Owing to the $1/\Rt$ factor multiplying the square bracket in Eq.~\eqref{E:A:jn}, the spherical Bessel function $j_n$ for even $n$ may thus be expressed as an expansion in even powers of $\Rt$. Consequently, since ${\RtI^2=\RtII^2}$ the spherical Bessel function $j_n$ for even $n$ is independent of the choice of the expression used for $\Rt$.  Also, since there are no odd powers of $\Rt$ in the expression for $j_n$ for even $n$, no branch cuts are introduced.  Thus, the phasor $U_{0,n}$ for even $n$ has no discontinuities. 

\section{Discontinuity in the Real Fields}\label{sec:DCfields}

We can express the phasor of Eq.~\eqref{E:fullPhasor} in the time domain via a Fourier transformation,

\begin{equation}\label{E:Fourier}
U_{0,n}(\mathbf{r},t)=\frac{1}{\sqrt{2\pi}}\int U_{0,n}(\mathbf{r},\omega)\exp(i\omega t)d\omega,
\end{equation}

\noindent the result of which is presented for arbitrary $n$ in Eq.~\eqref{E:A:fullPhasor} of Appendix A.  Recall that the frequency spectrum $f(\omega)$ of the pulse, defined in Eq.~\eqref{E:PoissonSpectrum}, introduces an overall phase factor $\exp(i\phi_0)$ in both the frequency-dependent and time-dependent phasors in Eqs.~\eqref{E:fullPhasor} and~\eqref{E:Fourier} respectively.  Therefore, changes in the initial phase $\phi_0$ can affect the occurrence of discontinuities in the real and imaginary components of the phasor.   

Figure~\ref{fig:phasors} shows explicitly the discontinuities in the time domain phasor for $n=3$ when using the choice $\Rt=\RtI$ for three values of the phase $\phi_0$. These plots were generated for a linearly polarized beam with spectral parameter $s=712$, beam waist ${w_0=2~\mu m}$, wavelength ${\lambda=800~nm}$, and Rayleigh length ${z_R=k w_0^2/2}$. As expected, no discontinuity is visible in the square modulus of the time domain phasor for any $\phi_0$. However, the discontinuity at ${z=0}$ is clearly visible in the real and/or imaginary parts of the phasor, depending upon the value of $\phi_0$ [cf. panels (c), (e), (f), and (h) of Fig.~\ref{fig:phasors}].

\begin{figure}[t]
	\centering
	\subfloat[][]{%
		\label{fig:m1Ez}%
		\includegraphics[height=2.5in]{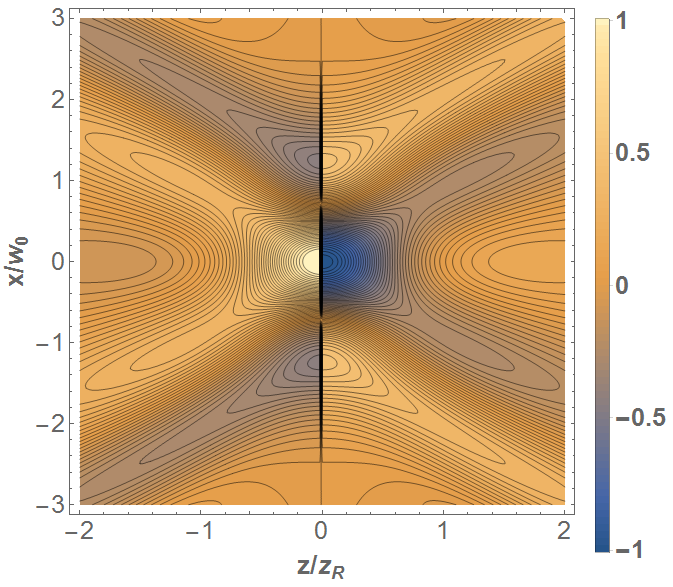}}%
	\\
	\subfloat[][]{%
		\label{fig:m3Ez}%
		\includegraphics[height=2.5in]{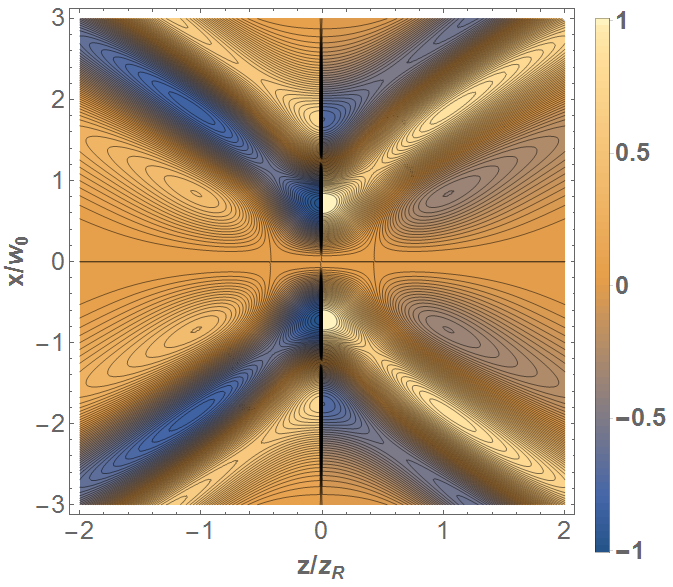}}%
	\caption[]{Discontinuities in the longitudinal fields $E_z$ [obtained using Eq.~\eqref{E:hertzfields}] across the beam waist $z=0$  for the odd OAM phasors $U_{0,n}$ for \subref{fig:m1Ez} ${n=1}$, and \subref{fig:m3Ez} ${n=3}$.  These fields were obtained using the choice $\Rt=\RtI$ for a phase ${\phi_0=\pi}$, $y=0$, and ${t=z/c}$. The amplitudes of the fields are normalized to unity. See text for discussion.}
	\label{fig:Ez}%
\end{figure}

In Figure~\ref{fig:Ez}, we plot the longitudinal fields $E_z$ [obtained using Eq.~\eqref{E:hertzfields}] for the odd OAM phasors $U_{0,n}(\mathbf{r},t)$ for $n=1$ and $n=3$ using the choice $\Rt=\RtI$ and an overall phase ${\phi_0=\pi}$.  This choice of the phase $\phi_0$ yields a discontinuity in the imaginary parts of each of the phasors, which in turn results in discontinuous fields $E_z$.  The corresponding transverse fields (not shown) are continuous across the beam waist at $z=0$. In general, for linearly polarized fields, our calculations show that discontinuities in the real part of the time domain phasor lead to discontinuities in the transverse components of $\mathbf{E}$ and $\mathbf{B}$ while discontinuities in the imaginary part of the time domain phasor lead to discontinuities in the longitudinal components of the fields.  When both the real and imaginary parts of the phasor have discontinuities, the problem appears in all real field components.

As for the case of linear polarization, other beam polarizations will also suffer discontinuous real fields for phasors calculated using the choice $\Rt=\RtI$.  These discontinuities originate in the phasor, which is polarization-independent.  The polarization only enters when computing the fields using the Hertz potentials, as Eq.~\eqref{E:Hertz} demonstrates for the case of linear polarization.  The discontinuities may occur in different field components, depending on the field polarization, but they will be present in the real fields nonetheless.

Although our focus in this paper is on solutions to the nonparaxial Helmholtz equation, a brief mention of the paraxial case is warranted.  In the paraxial limit of the phasor (cf.~Eq.~(5) of Ref.~\cite{April2010}), the terms ${P_n^n(\chi)}$ and $\Rt$ do not enter.  In fact, to lowest radial order the associated Laguerre polynomials in the paraxial phasor are unity.  Thus the real and imaginary parts of the paraxial phasor, by direct inspection, are simple oscillatory functions of $z$.  In this limit, therefore, the problem of discontinuities in the fields does not arise.

\section{Summary and Conclusions}\label{sec:conc}

In this work, we have shown (by examining the nonparaxial source/sink phasor) that for all odd OAM modes discontinuities arise across the entire beam waist when the choice $\Rt=\RtI$ is made for the complex spherical radius.  Whether these discontinuities lie in the real or imaginary parts of the phasor depends upon the overall phase $\phi_0$ of the laser pulse.  In turn, these phasor discontinuities result in nonphysical real electromagnetic field components calculated from the Hertz potentials.

As we have shown, these problems do not exist for even OAM modes.  Further, in the paraxial limit, the terms that cause discontinuous behavior are not present in the phasor expression.  Thus, real components of paraxial fields are free from discontinuities in the phasor that cause problems in the nonparaxial case.

Whether considering the fields of vortex beams in vacuum, or interacting with plasmas or other media, proper physical theoretical models are necessary.  As this work has shown, discontinuities in the nonparaxial source/sink phasor can be avoided completely by making the choice $\Rt=\RtII$ for the complex spherical radius.  Such a choice avoids discontinuities in the complex phasor for all OAM modes.

\section{Acknowledgments}

We gratefully acknowledge informative discussion with Alexandre April regarding his work.  Computational results were obtained using facilities at the Holland Computing Center of the University of Nebraska-Lincoln.  This work was supported in part by the U.S. Department of Energy, Office of Science, Basic Energy Sciences, under Award No. DE-FG02-96ER14646.

\onecolumngrid
\appendix
\section{Explicit Expression for the Phasor $U_{0,n}(\textbf{r}, t)$}
In this Appendix we present the result of carrying out the Fourier transform of the phasor ${U_{0,n}(\mathbf{r},\omega)}$ in Eq.~\eqref{E:fullPhasor}, which is obtained from Eqs.~(16)~\&~(17) of Ref.~\cite{April2010} for the phasor ${\tilde{U}^\sigma_{0,n}(\mathbf{r},\omega)}$ upon setting ${\sigma=e}$ and ${p=0}$ (and dropping the explicit notation of the parity ${\sigma=e}$ in our calculations).  In order to carry out the Fourier transform in Eq.~\eqref{E:Fourier}, one must expand the spherical Bessel function in Eq.~\eqref{E:fullPhasor} using Eq.~(10.1.8) of Ref.~\cite{Abramowitz}:

\begin{equation}\label{E:A:jn}
	j_n(k\Rt)=\frac{1}{k\Rt}\left[ P\left(n+\frac{1}{2},k\Rt\right)\sin\left(k\Rt-\frac{n\pi}{2}\right)\,+\,Q\left(n+\frac{1}{2},k\Rt\right)\cos\left(k\Rt-\frac{n\pi}{2}\right)\right]
\end{equation}

\noindent where

\begin{equation}\label{E:A:P}
	P\left(n+\frac{1}{2},k\Rt\right)=\sum_{m=0}^{\floor{n/2}}(-1)^m (2k\Rt)^{(-2m)}\frac{(n+2m)!}{(2m)!\,\Gamma(n-2m+1)}
\end{equation}

\noindent and

\begin{equation}\label{E:A:Q}
	Q\left(n+\frac{1}{2},k\Rt\right)=\sum_{m=0}^{\floor{(n-1)/2}}(-1)^m (2k\Rt)^{(-2m-1)}\frac{(n+2m+1)!}{(2m+1)!\,\Gamma(n-2m)}
\end{equation}

\subsection{Result for $U_{0,n}(\mathbf{r},t)$}

Expanding the trigonometric functions in Eq.~\eqref{E:A:jn} in terms of exponentials and replacing $k$ everywhere by ${k=\omega/c}$, one may carry out the Fourier transform in Eq.~\eqref{E:Fourier} by making repeated use of the integral representation of the gamma function (cf. Eq.~(6.1.1) of~\cite{Abramowitz}), i.e.,

\begin{equation}\label{E:A:gammaIntegral}
	\Gamma(\gamma+1)=\eta^{\gamma+1}\int_{0}^\infty \textup{d}\omega\,\omega^\gamma \exp(-\eta \omega)\quad,Re\,\eta>0
\end{equation}

\noindent The result for ${U_{0,n}(\mathbf{r},t)}$ is:

\begin{equation}\label{E:A:fullPhasor}
\begin{aligned}
	U_{0,n}(\mathbf{r},t)=C_n \cos(n\phi) P^n_n(\chi) \left\{ \sum_{m=0}^{\floor{n/2}}A(n,m)\left(\frac{c}{\Rt}\right)^{2m+1} \left[ (T_-)^{-(s+n/2-2m+1)}-(-1)^n(T_+)^{-(s+n/2-2m+1)} \right] \right.\\
	\left.+ \sum_{m=0}^{\floor{(n-1)/2}}D(n,m)\left(\frac{c}{\Rt}\right)^{2m+2} \left[ (T_-)^{-(s+n/2-2m)}+(-1)^n(T_+)^{-(s+n/2-2m)} \right] \right\}
\end{aligned}
\end{equation}

\noindent In Eq.~\eqref{E:A:fullPhasor} we have defined

\begin{equation}\label{E:A:A}
	A(n,m)\equiv\frac{i (-1)^{m+1} (n+2m)!}{(2m)!~\Gamma(n-2m+1)} \frac{\Gamma(s+n/2-2m+1)}{2^{(2m+1)}~\Gamma(s+1)} \left(\frac{s}{\omega_0}\right)^{(2m-n/2)}
\end{equation}

\begin{equation}\label{E:A:D}
	D(n,m)\equiv\frac{(-1)^m (n+2m+1)!}{(2m+1)!~\Gamma(n-2m)} \frac{\Gamma(s+n/2-2m)}{2^{(2m+2)}~\Gamma(s+1)} \left(\frac{s}{\omega_0}\right)^{(2m+1-n/2)}
\end{equation}

\vspace{0.1in}
\noindent where $s$ and $\omega_0$ are defined in the text below Eq.~\eqref{E:PoissonSpectrum},

\begin{equation}\label{E:A:Cn}
	C_n\equiv\exp[i~(\phi_0-n\pi/2)] \left(\frac{a}{c}\right)^{(1+n/2)} \frac{2^{(1-n/2)}}{(2n-1)!!}
\end{equation}

\noindent and

\begin{equation}
	T_\pm\equiv~1-\frac{i \omega_0 t}{s}+\frac{a \omega_0}{c s} \pm \frac{i \omega_0 \Rt}{c s}
\end{equation}

\subsection{Result for $U_{0,1}(\mathbf{r},t)$}

Setting ${n=1}$ in Eq.~\eqref{E:A:fullPhasor}, we have

\begin{equation}\label{E:A:C1}
	C_1=\exp[i~(\phi_0-\pi/2)] \sqrt{2} \left(\frac{a}{c}\right)^{3/2}
\end{equation}

\begin{equation}\label{E:A:A10}
	A(1,0)= -i~\frac{\Gamma(s+3/2)}{2~\Gamma(s+1)} \left(\frac{s}{\omega_0}\right)^{-1/2}
\end{equation}

\begin{equation}\label{E:A:D10}
	D(1,0)=\frac{\Gamma(s+1/2)}{2~\Gamma(s+1)} \left(\frac{s}{\omega_0}\right)^{1/2}
\end{equation}

\noindent Hence, 

\begin{equation}\label{E:A:U01}
\begin{aligned}
	U_{0,1}(\mathbf{r},t)=C_1 \cos(\phi) P^1_1(\chi) \left\{ A(1,0)\left(\frac{c}{\Rt}\right) \left[ (T_-)^{-(s+3/2)}+(T_+)^{-(s+3/2)}\right] +\right.\\
	+~\left. D(1,0)\left(\frac{c}{\Rt}\right)^2 \left[ (T_-)^{-(s+1/2)}-(T_+)^{-(s+1/2)} \right] \right\}
\end{aligned}
\end{equation}

\twocolumngrid
\nocite{apsrev41Control}

\end{document}